\documentclass[lettersize,journal]{IEEEtran}
\usepackage[utf8]{inputenc}
\usepackage[T1]{fontenc}
\usepackage{amsmath,amsfonts}
\usepackage{algorithmic}
\usepackage{algorithm}
\usepackage{array}
\usepackage[caption=false,font=normalsize,labelfont=sf,textfont=sf]{subfig}
\usepackage{textcomp}
\usepackage{stfloats}
\usepackage{url}
\usepackage{verbatim}
\usepackage{graphicx}
\usepackage{cite}
\usepackage{siunitx}
\usepackage{booktabs}
\usepackage{multirow}
\usepackage{enumitem}
\setlist{nosep,leftmargin=*}
\usepackage{pgfplots}
\usepgfplotslibrary{statistics}
\pgfplotsset{compat=1.18, every axis plot/.append style={thick}}
\usepackage{tikz}
\usetikzlibrary{arrows.meta, decorations.pathreplacing}

\definecolor{crandom}{RGB}{58,82,139}
\definecolor{coracle}{RGB}{94,201,97}
\definecolor{csurrogate}{RGB}{225,100,98}
\definecolor{cdcf}{RGB}{89,161,79}
\definecolor{cbaseline}{RGB}{156,102,141}
\definecolor{chmab}{RGB}{98,163,158}
\definecolor{cfoptimal}{RGB}{242,142,43}
\definecolor{ctoptimal}{RGB}{142,43,242}

\begin{document}

\onecolumn{

\thispagestyle{empty}
\setcounter{page}{0}

\large
\noindent Please cite this paper as: \newline
M. Wojnar, "Sample-and-Score: Surrogate-Guided Coordinated Spatial Reuse in IEEE 802.11 Networks," in IEEE Communications Letters, doi: 10.1109/LCOMM.2026.3746794.

\normalsize
\vspace{1cm}
\begin{verbatim}

@ARTICLE{wojnar2026sample,
  author={Wojnar, Maksymilian},
  journal={IEEE Communications Letters}, 
  title={{Sample-and-Score: Surrogate-Guided Coordinated 
  Spatial Reuse in IEEE 802.11 Networks}}, 
  year={2026},
  doi={10.1109/LCOMM.2026.3746794}
}

\end{verbatim}

}
\twocolumn
\clearpage

\bstctlcite{IEEEexample:BSTcontrol}

\title{Sample-and-Score: Surrogate-Guided Coordinated Spatial Reuse in IEEE 802.11 Networks}

\author{Maksymilian Wojnar
\thanks{Maksymilian Wojnar is with the AGH University of Krakow, Poland,\break email: maksymilian.wojnar@agh.edu.pl. This research was funded by the National Science Centre, Poland (DEC-2020/39/I/ST7/01457). We gratefully acknowledge Polish high-performance computing infrastructure PLGrid (HPC Center: ACK Cyfronet AGH) for providing computer facilities and support within computational grant no. PLG/2025/018718. }} 

\maketitle

\begin{abstract}
Coordinated spatial reuse (Co-SR) in IEEE 802.11bn increases area throughput through simultaneous transmissions, yet the combinatorial configuration space makes real-time optimization challenging. We propose a sample-and-score approach: draw random Co-SR configurations, score each with a surrogate model, and select the best. The surrogate is trained offline and deployed without online fine-tuning or sequential exploration, yet performs close to multi-armed bandit and heuristic baselines that need multiple interaction steps or extensive measurements. On unseen topologies with 2--16 access points, it recovers 52--77\% of a perfect surrogate's throughput. A perfect sample-and-score approach reaches 95\% of the upper bound throughput with 8192 candidates.
\end{abstract}


\section{Introduction}
\label{sec:introduction}

Wi-Fi networks are shifting from competitive channel access to collaborative coordination among access points (APs). With multi-AP coordination (MAPC), a defining feature of IEEE~802.11bn (Wi-Fi~8), a pair of APs coordinate their transmissions for improved spectral efficiency~\cite{ning2025survey}. Coordinated spatial reuse (Co-SR), a particularly promising MAPC scheme, allows multiple APs to transmit simultaneously on the same channel by jointly tuning transmit power levels. This can significantly increase aggregate throughput in dense deployments where overlapping basic service sets are inevitable. However, this comes at a cost: each simultaneous transmission adds interference at other receivers, so transmit power, the served station (STA), and modulation and coding scheme (MCS) must be optimized jointly to maintain a sufficiently high signal-to-interference-and-noise ratio (SINR) at every receiver. Striking this balance is a delicate optimization problem.

Orchestrating Co-SR transmissions in real time is challenging due to the combinatorial explosion in the number of possible configurations. Centralized heuristics driven by periodic received signal strength indicator (RSSI) reports~\cite{haxhibeqiri2024coordinated, nunez2023group} and convex-optimization-based methods~\cite{zhu2025two} work in small topologies but become intractable as the network grows. Online learning methods such as multi-armed bandits (MABs) adapt without channel knowledge~\cite{wojnar2025ieee, wojnar2025csr}, yet their convergence time increases with topology size. Deep reinforcement learning approaches~\cite{du2025deep, jung2026coordinated} select from a fixed catalog of pre-formed Co-SR groups, requiring action-space regeneration and retraining for unseen deployments. A common limitation across these methods is that they decouple or fix MCS selection; although MCS negotiation is not formally part of the Co-SR specification, it critically determines achievable rates.

\begin{figure}[t]
    \centering
    \input{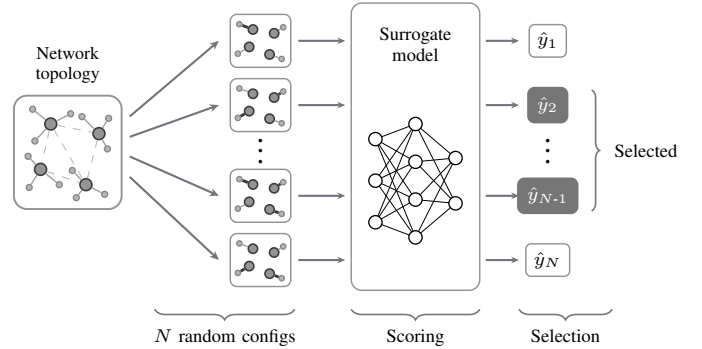}
    \caption{Overview of the proposed surrogate-based Co-SR optimization pipeline: random configurations are encoded as graphs, scored by the surrogate, and the best ones are selected for execution.}
    \label{fig:pipeline}
\end{figure}

In this letter, we propose a simpler alternative: \emph{draw $N$ random Co-SR configurations, score each with a surrogate model, and select the best ones}. Fig.~\ref{fig:pipeline} illustrates the proposed pipeline. Surrogate-based optimization is a well-established paradigm~\cite{forrester2008engineering}. Our contribution lies in applying it to the Co-SR configuration problem with a graph-structured representation that generalizes across topologies and scales to groups of up to 16~APs, extending beyond the pair-wise scope of the current Wi-Fi~8 specification. The surrogate is trained offline and replaces simulation at test time, requiring only a single forward pass and no convergence period, unlike online methods that need hundreds of interactions to reach the target performance. Controller-side planners increasingly rely on digital-twin models to evaluate configurations before deployment~\cite{sari2026uwb}. Our surrogate plays a similar role as a fast, learned twin, replacing analytical planning with sampling-and-scoring. The source code is publicly available.\footnote{\url{https://github.com/ml4wifi-devs/mapc-surrogate}}

\section{Problem Formulation}
\label{sec:problem}

A Co-SR configuration specifies: (i)~which APs transmit simultaneously, (ii)~which STA each active AP serves, (iii)~the transmission power level, and (iv)~the MCS index for each active AP. The configuration space grows combinatorially with the number of APs and STAs. For a network with $|\mathcal{A}|$~APs, each serving up to $|\mathcal{S}_a|$~STAs associated with it, the number of possible configurations is
\begin{equation}
    \prod_{a \in \mathcal{A}} \bigl(1 + |\mathcal{S}_a| \cdot N_\text{MCS} \cdot N_\text{TP}\bigr),
    \label{eq:config_space}
\end{equation}
where $N_\text{MCS}$ and $N_\text{TP}$ are the number of MCS indices and transmission power levels, respectively, and $+1$ accounts for the option of an AP not transmitting. With $N_\text{MCS} = 14$, $N_\text{TP} = 4$, and $|\mathcal{S}_a| = 4$, a network of just 5~APs already faces over $10^{11}$~possible configurations, making exhaustive search infeasible at per-transmission opportunity (TXOP) timescales.

We evaluate our approach using \emph{mapc-sim}\footnote{\url{https://github.com/ml4wifi-devs/mapc-sim}}~\cite{wojnar2025ieee}, a custom open-source Monte Carlo MAPC simulator, as currently no full-stack simulator (e.g., ns-3) supports Co-SR. Our simulator computes path loss according to the TGax enterprise model~\cite{tgax}, derives the SINR from the set of concurrent transmissions, and estimates the probability of successful aggregate MAC protocol data unit (A-MPDU) sub-frame reception using a binomial model based on the selected MCS. Nakagami-m fading is introduced to model small-scale channel variation. The optimization objective is the network throughput. In line with the Wi-Fi~8 Co-SR design, we consider downlink (AP-to-STA) transmissions only; uplink transmissions follow DCF and are excluded from evaluation. Every Co-SR transmission is preceded by our proposed MAPC signaling: the sharing AP sends a Co-SR invite, the invited APs reply simultaneously over OFDMA, and a trigger frame confirms the agreed parameters, with the acknowledgments likewise sent over OFDMA. We account for the backoff procedure, the interframe spaces, and the signaling overhead using the frame lengths and timings of~\cite{wilhelmi2026tutorial}. We evaluate under full-buffer traffic, the worst-case load for throughput maximization. Table~\ref{tab:sim_params} summarizes the simulation parameters.

\begin{table}
    \caption{Simulation parameters.}
    \footnotesize
    \label{tab:sim_params}
    \centering
    \begin{tabular}{@{\hskip 4pt}ll@{\hskip 4pt}}
        \toprule
        \textbf{Parameter}         & \textbf{Value}            \\ \midrule
        Band                       & 5 GHz                     \\
        PHY                        & IEEE 802.11be             \\
        Channel width              & 80 MHz                    \\
        Spatial streams            & 1, SISO                   \\
        Path loss model            & TGax w/ Nakagami fading   \\
        Nakagami-m parameter       & 1.5                       \\
        CCA threshold              & \SI{-82}{dBm}             \\
        Frame aggregation          & Enabled (A-MPDU)          \\
        A-MPDU sub-frame size      & 1500 B                    \\
        TXOP duration              & \SI{5.484}{\milli\second} \\
        Traffic                    & Downlink, full buffer     \\ \bottomrule
    \end{tabular}
\end{table}

\section{Proposed Method}
\label{sec:method}

We model the Wi-Fi network as an undirected graph $\mathcal{G} = (\mathcal{V}, \mathcal{E})$, where $\mathcal{V} = \mathcal{A} \cup \mathcal{S}$ comprises APs and STAs. The edge set $\mathcal{E}$ contains AP--STA \emph{transmission links} (one per association) and AP--AP \emph{interference links} (between APs whose mutual RSSI exceeds the clear channel assessment (CCA) threshold). Each edge carries a four-attribute feature vector:
\begin{itemize}
    \item \textit{rssi} $\in \mathbb{R}$: RSSI, standardized as $\tilde{r}_s = (r_s - \mu_\text{RSSI}) / \sigma_\text{RSSI}$ with $\mu_\text{RSSI}$ and $\sigma_\text{RSSI}$ computed from the training set,
    \item \textit{selected} $\in \{\text{True}, \text{False}, \text{N/A}\}$: whether the link is chosen for transmission,
    \item \textit{mcs} $\in \{0, \ldots, 13, \text{N/A}\}$: MCS index of the selected link,
    \item \textit{tx\_power} $\in \{16, 13, 10, 7, \text{N/A}\}~\si{dBm}$: transmission power level.
\end{itemize}
Categorical attributes are one-hot encoded while RSSI remains a continuous scalar, yielding a 24-dimensional edge feature vector. Edges are bidirectional. The N/A class handles inapplicable attributes (e.g., MCS on interference links). A central controller gathers these measurements, scores the candidate configurations with the surrogate, and sends the selected ones back to the APs. This centralized architecture targets managed Wi-Fi deployments, where coordination among multiple APs is typically available through a network controller.

During a Co-SR transmission, the interference at a given STA depends on its channel to every other active AP. Capturing all of these links directly would require $|\mathcal{A}| \times |\mathcal{S}|$ RSSI measurements, far more than the network gathers in normal operation. We instead retain only the AP--AP and associated AP--STA links, which are available at little cost: inter-AP RSSI is obtained by listening to beacons, and AP-to-STA RSSI is derived from acknowledgment frames through channel reciprocity. The surrogate then infers the residual cross-interference through message passing, a deliberate approximation that trades exactness for a lower signaling overhead.

The surrogate model takes a graph-encoded Co-SR configuration and predicts the data rate. The model is trained offline on simulated data and replaces costly network simulation during candidate evaluation. The neural network architecture follows the graph network framework~\cite{battaglia2018relational} with transformer-based~\cite{vaswani2017attention} update functions (3 GNN layers with residual connections, a 3-layer transformer, a hidden dimension of 64, 4~attention heads). Attention is masked, so that graphs within a batch are scored independently. 

The output layer follows the mixture density network (MDN) framework~\cite{bishop1994mdn} with 4~Gaussian components. MDN is preferred over a standard mean squared error (MSE) regression as the stochastic data rate is better captured by a mixture and in our experiments MSE overfits where MDN does not. Each component is parametrized by a weight $\hat{w}_k = \text{softmax}(\boldsymbol{\alpha})_k$, where $\boldsymbol{\alpha}$ is the vector of output logits, a mean $\hat{\mu}_k$, and a scale $\hat{\sigma}_k$. The model is trained by minimizing the negative log-likelihood of the standardized data rate $\tilde{y} = (y - \mu_y) / \sigma_y$ under the predicted Gaussian mixture:
\begin{equation}
    \mathcal{L} = -\log \sum_{k=1}^{m} w_k \cdot \mathcal{N}(\tilde{y} \mid \hat{\mu}_k, \hat{\sigma}_k^2).
    \label{eq:mdn_loss}
\end{equation}
The model is trained with AdamW~\cite{loshchilov2017decoupled} (learning rate $1.5 \times 10^{-3}$, weight decay $0.01$, gradient clipping) under a cosine one-cycle schedule for 20,000~steps with batch size~16. On a hold-out validation set, the surrogate achieves a Spearman correlation of $\rho = 0.87$ and $R^2 = 0.70$.

The training set is generated from 11~parameterized scenario types with 2--10~APs and up to 50~STAs. For each of the 1,000~randomly sampled realizations, 50~Co-SR configurations are drawn: 40 with fully random parameters and 10 with simulator-computed ideal MCS to increase the high-rate region coverage, yielding 50,000~training examples.

Given a network topology, its associations, and the collected RSSI values, the inference pipeline proceeds as follows:
\begin{enumerate}
    \item \textbf{Random sampling.} Draw $N$ random Co-SR configurations by selecting $k \sim \text{Uniform}(1, |\mathcal{A}|)$ active APs, one STA per active AP, and random MCS and transmission power settings. Each configuration is encoded as a graph $\mathcal{G}_i$.
    \item \textbf{Surrogate scoring.} Batch the $N$ graphs and compute the expected data rate for each candidate as $\hat{y}_i = \sum_k w_k \hat{\mu}_k$.
    \item \textbf{Selection.} Select configurations to apply in the network according to the chosen strategy.
\end{enumerate}

The selection step introduces an additional degree of freedom that can be tailored to the optimization objective. The proposed method is therefore best understood as a scheduling aid: it supplies a pre-computed catalog of high-quality Co-SR groups that a network controller can apply according to its own policy (e.g., factoring in latency or fairness), rather than acting as a standalone scheduler. The simplest strategy, adopted in our experiments, is \emph{top-$k$}: retain the $k$ highest-scoring configurations and apply them in subsequent TXOPs in a round-robin fashion.

By design, the method operates on a snapshot of the current network state and is therefore directly applicable to static topologies. For dynamic environments, the catalog is regenerated periodically or adaptively when a throughput drop or association change is detected.

\section{Performance Evaluation}
\label{sec:performance}

\begin{figure}[t]
    \centering
    \begin{tikzpicture}
\begin{axis}[
  width=\columnwidth,
  height=0.62\columnwidth,
  xlabel={Number of candidates},
  ylabel={Throughput [\si{\mega\bit\per\second}]},
  ymin=0, ymax=1600,
  ytick distance=400,
  xmode=log,
  xmin=0.7, xmax=12000,
  log basis x=2,
  xtick={1,2,4,8,16,32,64,128,256,512,1024,2048,4096,8192},
  xticklabels={$2^0$,$2^1$,$2^2$,$2^3$,$2^4$,$2^5$,$2^6$,$2^7$,$2^8$,$2^9$,$2^{10}$,$2^{11}$,$2^{12}$,$2^{13}$},
  grid=both,
  major grid style={line width=0.2pt, draw=gray!40},
  minor grid style={line width=0.1pt, draw=gray!20},
  tick label style={font=\small},
  label style={font=\small},
  every axis plot/.append style={thick, mark size=1.6pt},
  legend style={
    at={(0.5,0.98)},
    anchor=north,
    draw=black,
    fill=white,
    fill opacity=0.9,
    text opacity=1,
    font=\scriptsize,
    rounded corners=2pt,
    legend columns=3
  },
  legend cell align=left,
]

\addlegendimage{color=csurrogate, solid, thick, mark=*}
\addlegendentry{Surrogate (ours)}
\addlegendimage{color=cdcf, densely dashed, thick}
\addlegendentry{DCF}
\addlegendimage{color=cfoptimal, dashed, thick}
\addlegendentry{F-optimal}
\addlegendimage{color=coracle, solid, thick, mark=triangle*}
\addlegendentry{Perfect surrogate (ours)}
\addlegendimage{color=chmab, solid, thick}
\addlegendentry{H-MAB}
\addlegendimage{color=ctoptimal, densely dashdotted, thick}
\addlegendentry{T-optimal}
\addlegendimage{color=crandom, dotted, thick}
\addlegendentry{Random}
\addlegendimage{color=cbaseline, dash dot, thick}
\addlegendentry{SR groups}

\addplot[color=crandom, fill=crandom, fill opacity=0.15, draw=none, forget plot]
  coordinates {(0.7,85.42) (12000,85.42) (12000,156.52) (0.7,156.52)} \closedcycle;
\addplot[color=crandom, dotted, thick, mark=none, forget plot]
  coordinates {(0.7,120.97) (12000,120.97)};

\addplot[color=cdcf, fill=cdcf, fill opacity=0.15, draw=none, forget plot]
  coordinates {(0.7,566.95) (12000,566.95) (12000,568.83) (0.7,568.83)} \closedcycle;
\addplot[color=cdcf, densely dashed, thick, mark=none, forget plot]
  coordinates {(0.7,567.89) (12000,567.89)};

\addplot[color=chmab, fill=chmab, fill opacity=0.15, draw=none, forget plot]
  coordinates {(0.7,733.78) (12000,733.78) (12000,741.09) (0.7,741.09)} \closedcycle;
\addplot[color=chmab, solid, thick, mark=none, forget plot]
  coordinates {(0.7,737.43) (12000,737.43)};

\addplot[color=cbaseline, fill=cbaseline, fill opacity=0.15, draw=none, forget plot]
  coordinates {(0.7,704.17) (12000,704.17) (12000,775.93) (0.7,775.93)} \closedcycle;
\addplot[color=cbaseline, dash dot, thick, mark=none, forget plot]
  coordinates {(0.7,740.05) (12000,740.05)};

\addplot[color=cfoptimal, dashed, thick, mark=none, forget plot]
  coordinates {(0.7,763.25) (12000,763.25)};
\addplot[color=ctoptimal, densely dashdotted, thick, mark=none, forget plot]
  coordinates {(0.7,1025.02) (12000,1025.02)};

\addplot[color=coracle, solid, mark=triangle*, forget plot, error bars/.cd, y dir=both, y explicit]
  coordinates {
    (1,155.76) +- (0,40.76)
    (2,260.74) +- (0,42.65)
    (4,355.52) +- (0,40.23)
    (8,452.49) +- (0,36.33)
    (16,569.70) +- (0,24.65)
    (32,637.49) +- (0,20.58)
    (64,674.65) +- (0,17.12)
    (128,721.80) +- (0,19.22)
    (256,754.15) +- (0,19.05)
    (512,819.58) +- (0,20.04)
    (1024,864.49) +- (0,16.85)
    (2048,908.06) +- (0,14.26)
    (4096,941.07) +- (0,13.01)
    (8192,972.88) +- (0,10.56)
  };

\addplot[color=csurrogate, solid, mark=*, forget plot, error bars/.cd, y dir=both, y explicit]
  coordinates {
    (1,120.97) +- (0,35.55)
    (2,214.77) +- (0,44.37)
    (4,257.99) +- (0,44.95)
    (8,343.03) +- (0,40.93)
    (16,473.67) +- (0,32.16)
    (32,498.30) +- (0,30.51)
    (64,530.34) +- (0,23.29)
    (128,557.45) +- (0,19.52)
    (256,588.53) +- (0,19.94)
    (512,609.91) +- (0,27.12)
    (1024,656.01) +- (0,14.82)
    (2048,672.02) +- (0,10.17)
    (4096,677.93) +- (0,5.79)
    (8192,671.41) +- (0,17.73)
  };

\end{axis}
\end{tikzpicture}
    \caption{Throughput vs.\ number of candidates $N$ in the residential scenario ($2 \times 2$ apartments) with top-1 selection. Error bars denote 95\% confidence intervals of the mean.}
    \label{fig:sweep}
\end{figure}
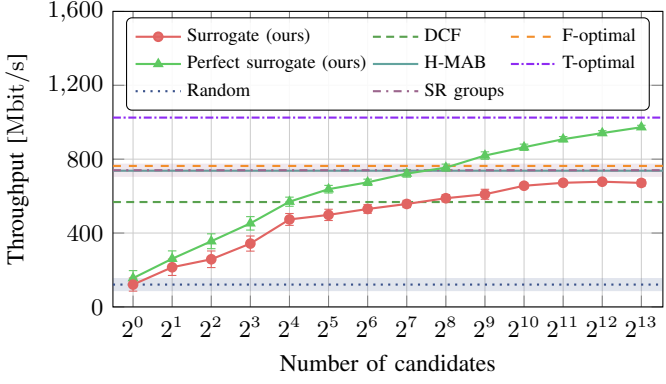

The \emph{residential} scenario follows the TGax enterprise model~\cite{tgax}: a $2 \times 2$ grid of \SI{10}{\meter}-wide apartments, each containing one AP and four randomly placed STAs, with walls providing additional attenuation. The \emph{open-space} scenario places 2--16 APs at random within a $\SI{75}{\meter} \times \SI{75}{\meter}$ area without walls, with four STAs per AP normally distributed around their associated AP (standard deviation of $\SI{5}{\meter}$).

We compare the following methods:
\begin{itemize}
    \item \textbf{Surrogate (ours):} draws $N$ candidates, scores each with the trained surrogate, and selects the best ones.
    \item \textbf{Perfect surrogate (ours):} deterministically evaluates the expected data rate of each of the $N$ candidates and selects the best one; this is the upper bound for the sample-and-score approach.
    \item \textbf{Random:} picks one random Co-SR configuration.
    \item \textbf{DCF:} standard 802.11 distributed coordination function.\footnote{\url{https://github.com/ml4wifi-devs/mapc-dcf}}
    \item \textbf{SR groups:} the at-most-$K$ SR-compatible group-formation heuristic~\cite{nunez2023group}: APs are greedily grouped so that the estimated SINR stays above a threshold ($\gamma = \SI{20}{dB}$, $K = 3$, fixed transmission power).
    \item \textbf{H-MAB:} hierarchical MAB~\cite{wojnar2025ieee} with three levels of agents that sequentially select active APs, recipient STAs, and transmission power levels.
    \item \textbf{F-optimal:} the fair upper bound~\cite{wojnar2025csr}, maximizing the minimum per-STA throughput, enforcing perfectly equal shares. A globally optimal reference, not a runtime scheduler.
    \item \textbf{T-optimal:} the throughput upper bound~\cite{wojnar2025csr}, maximizing total network throughput. A globally optimal reference, not a runtime scheduler.
\end{itemize}
All methods are averaged over 64~independent runs. For H-MAB, only the last 20\% of the 2000-step runs are used. In all experiments, the surrogate methods jointly sample MCS with all configuration parameters, whereas DCF and H-MAB use ideal MCS. SR groups uses its own MCS heuristic, while F/T-optimal jointly optimize MCS. We retain the default MCS selection strategy for each baseline. This choice favors the baselines, as our sample-and-score approach must instead identify a suitable MCS solely from randomly sampled candidates.

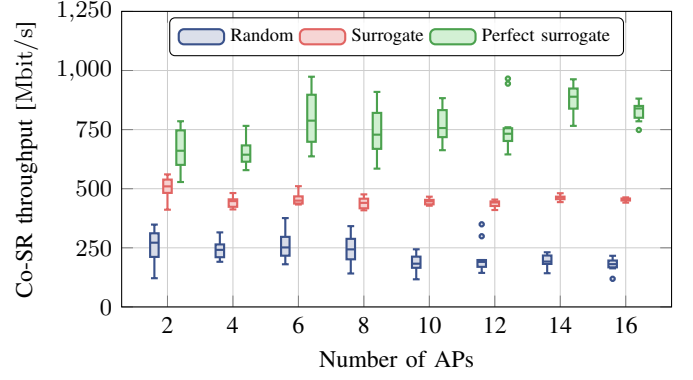
\begin{figure}[t]
    \centering
    \begin{tikzpicture}
\pgfmathsetmacro{\bw}{0.13}
\pgfmathsetmacro{\oA}{-0.20} 
\pgfmathsetmacro{\oC}{0.0}   
\pgfmathsetmacro{\oD}{0.20}  

\begin{axis}[
  width=\columnwidth,
  height=0.62\columnwidth,
  xlabel={Number of APs},
  ylabel={Co-SR throughput [\si{\mega\bit\per\second}]},
  ymin=0, ymax=1250,
  ytick distance=250,
  xtick={1,2,3,4,5,6,7,8},
  xticklabels={2,4,6,8,10,12,14,16},
  xmin=0.3, xmax=8.7,
  grid=both,
  major grid style={line width=0.2pt, draw=gray!40},
  minor grid style={line width=0.1pt, draw=gray!20},
  tick label style={font=\small},
  label style={font=\small},
  legend style={
    at={(0.5,0.98)},
    anchor=north,
    draw=black,
    fill=white,
    fill opacity=0.9,
    text opacity=1,
    font=\scriptsize,
    rounded corners=2pt,
    legend columns=3,
  },
  legend cell align=left,
  boxplot/draw direction=y,
]

\addlegendimage{area legend, draw=crandom, fill=crandom!20}
\addlegendentry{Random}
\addlegendimage{area legend, draw=csurrogate, fill=csurrogate!30}
\addlegendentry{Surrogate}
\addlegendimage{area legend, draw=coracle!80!black, fill=coracle!30}
\addlegendentry{Perfect surrogate}

\addplot+[boxplot prepared={draw position={1+\oA},
  lower whisker=121.4, lower quartile=211.6, median=271.7,
  upper quartile=311.5, upper whisker=348.3},
  draw=crandom, fill=crandom!20, boxplot/box extend=\bw, forget plot, mark=*, mark options={scale=0.4}]
  coordinates {};
\addplot+[boxplot prepared={draw position={2+\oA},
  lower whisker=190.9, lower quartile=210.3, median=241.4,
  upper quartile=264.4, upper whisker=315.3},
  draw=crandom, fill=crandom!20, boxplot/box extend=\bw, forget plot, mark=*, mark options={scale=0.4}]
  coordinates {};
\addplot+[boxplot prepared={draw position={3+\oA},
  lower whisker=180.4, lower quartile=216.7, median=251.6,
  upper quartile=296.4, upper whisker=375.8},
  draw=crandom, fill=crandom!20, boxplot/box extend=\bw, forget plot, mark=*, mark options={scale=0.4}]
  coordinates {};
\addplot+[boxplot prepared={draw position={4+\oA},
  lower whisker=141.4, lower quartile=201.4, median=243.5,
  upper quartile=288.0, upper whisker=341.7},
  draw=crandom, fill=crandom!20, boxplot/box extend=\bw, forget plot, mark=*, mark options={scale=0.4}]
  coordinates {};
\addplot+[boxplot prepared={draw position={5+\oA},
  lower whisker=116.8, lower quartile=165.5, median=182.9,
  upper quartile=212.7, upper whisker=243.8},
  draw=crandom, fill=crandom!20, boxplot/box extend=\bw, forget plot, mark=*, mark options={scale=0.4}]
  coordinates {};
\addplot+[boxplot prepared={draw position={6+\oA},
  lower whisker=143.9, lower quartile=169.6, median=190.5,
  upper quartile=198.3, upper whisker=198.4},
  draw=crandom, fill=crandom!20, boxplot/box extend=\bw, forget plot, mark=*, mark options={scale=0.4}]
  coordinates {(0,299.0)(0,349.5)};
\addplot+[boxplot prepared={draw position={7+\oA},
  lower whisker=142.5, lower quartile=181.8, median=191.9,
  upper quartile=217.5, upper whisker=231.2},
  draw=crandom, fill=crandom!20, boxplot/box extend=\bw, forget plot, mark=*, mark options={scale=0.4}]
  coordinates {};
\addplot+[boxplot prepared={draw position={8+\oA},
  lower whisker=163.2, lower quartile=167.4, median=180.9,
  upper quartile=196.3, upper whisker=216.3},
  draw=crandom, fill=crandom!20, boxplot/box extend=\bw, forget plot, mark=*, mark options={scale=0.4}]
  coordinates {(0,118.9)};

\addplot+[boxplot prepared={draw position={1+\oC},
  lower whisker=411.2, lower quartile=482.3, median=509.9,
  upper quartile=538.6, upper whisker=560.6},
  draw=csurrogate, fill=csurrogate!30, boxplot/box extend=\bw, forget plot, mark=*, mark options={scale=0.4}]
  coordinates {};
\addplot+[boxplot prepared={draw position={2+\oC},
  lower whisker=412.4, lower quartile=423.5, median=447.6,
  upper quartile=455.1, upper whisker=481.7},
  draw=csurrogate, fill=csurrogate!30, boxplot/box extend=\bw, forget plot, mark=*, mark options={scale=0.4}]
  coordinates {};
\addplot+[boxplot prepared={draw position={3+\oC},
  lower whisker=433.2, lower quartile=437.3, median=449.6,
  upper quartile=467.6, upper whisker=510.7},
  draw=csurrogate, fill=csurrogate!30, boxplot/box extend=\bw, forget plot, mark=*, mark options={scale=0.4}]
  coordinates {};
\addplot+[boxplot prepared={draw position={4+\oC},
  lower whisker=408.7, lower quartile=417.2, median=440.1,
  upper quartile=457.8, upper whisker=476.1},
  draw=csurrogate, fill=csurrogate!30, boxplot/box extend=\bw, forget plot, mark=*, mark options={scale=0.4}]
  coordinates {};
\addplot+[boxplot prepared={draw position={5+\oC},
  lower whisker=428.1, lower quartile=433.9, median=446.9,
  upper quartile=452.9, upper whisker=466.3},
  draw=csurrogate, fill=csurrogate!30, boxplot/box extend=\bw, forget plot, mark=*, mark options={scale=0.4}]
  coordinates {};
\addplot+[boxplot prepared={draw position={6+\oC},
  lower whisker=409.9, lower quartile=426.9, median=441.9,
  upper quartile=446.2, upper whisker=454.1},
  draw=csurrogate, fill=csurrogate!30, boxplot/box extend=\bw, forget plot, mark=*, mark options={scale=0.4}]
  coordinates {};
\addplot+[boxplot prepared={draw position={7+\oC},
  lower whisker=443.3, lower quartile=455.6, median=463.5,
  upper quartile=466.5, upper whisker=480.9},
  draw=csurrogate, fill=csurrogate!30, boxplot/box extend=\bw, forget plot, mark=*, mark options={scale=0.4}]
  coordinates {};
\addplot+[boxplot prepared={draw position={8+\oC},
  lower whisker=441.1, lower quartile=450.1, median=454.4,
  upper quartile=458.6, upper whisker=463.4},
  draw=csurrogate, fill=csurrogate!30, boxplot/box extend=\bw, forget plot, mark=*, mark options={scale=0.4}]
  coordinates {};

\addplot+[boxplot prepared={draw position={1+\oD},
  lower whisker=528.3, lower quartile=600.8, median=660.6,
  upper quartile=746.8, upper whisker=785.4},
  draw=coracle!80!black, fill=coracle!30, boxplot/box extend=\bw, forget plot, mark=*, mark options={scale=0.4}]
  coordinates {};
\addplot+[boxplot prepared={draw position={2+\oD},
  lower whisker=578.5, lower quartile=614.2, median=643.9,
  upper quartile=683.2, upper whisker=765.8},
  draw=coracle!80!black, fill=coracle!30, boxplot/box extend=\bw, forget plot, mark=*, mark options={scale=0.4}]
  coordinates {};
\addplot+[boxplot prepared={draw position={3+\oD},
  lower whisker=636.8, lower quartile=698.5, median=787.9,
  upper quartile=897.5, upper whisker=973.7},
  draw=coracle!80!black, fill=coracle!30, boxplot/box extend=\bw, forget plot, mark=*, mark options={scale=0.4}]
  coordinates {};
\addplot+[boxplot prepared={draw position={4+\oD},
  lower whisker=584.9, lower quartile=668.6, median=728.6,
  upper quartile=820.7, upper whisker=909.5},
  draw=coracle!80!black, fill=coracle!30, boxplot/box extend=\bw, forget plot, mark=*, mark options={scale=0.4}]
  coordinates {};
\addplot+[boxplot prepared={draw position={5+\oD},
  lower whisker=662.9, lower quartile=718.2, median=756.6,
  upper quartile=832.5, upper whisker=883.1},
  draw=coracle!80!black, fill=coracle!30, boxplot/box extend=\bw, forget plot, mark=*, mark options={scale=0.4}]
  coordinates {};
\addplot+[boxplot prepared={draw position={6+\oD},
  lower whisker=645.0, lower quartile=701.9, median=732.5,
  upper quartile=755.7, upper whisker=760.0},
  draw=coracle!80!black, fill=coracle!30, boxplot/box extend=\bw, forget plot, mark=*, mark options={scale=0.4}]
  coordinates {(0,944.5)(0,965.7)};
\addplot+[boxplot prepared={draw position={7+\oD},
  lower whisker=765.8, lower quartile=838.7, median=889.1,
  upper quartile=923.8, upper whisker=962.9},
  draw=coracle!80!black, fill=coracle!30, boxplot/box extend=\bw, forget plot, mark=*, mark options={scale=0.4}]
  coordinates {};
\addplot+[boxplot prepared={draw position={8+\oD},
  lower whisker=785.1, lower quartile=800.1, median=839.4,
  upper quartile=850.1, upper whisker=881.2},
  draw=coracle!80!black, fill=coracle!30, boxplot/box extend=\bw, forget plot, mark=*, mark options={scale=0.4}] coordinates {(0,748.4)};

\end{axis}
\end{tikzpicture}
    \caption{Co-SR throughput across open-space scenarios with 2--16~APs (top-4 selection, number of candidates $N$ ranging from 64 to 512).}
    \label{fig:boxplot}
\end{figure}

Fig.~\ref{fig:sweep} evaluates the residential scenario where we assess how performance scales with the number of sampled candidates $N$. The throughput of the perfect surrogate increases from \SI{156}{\mega\bit\per\second} at $N=1$ to \SI{973}{\mega\bit\per\second} at $N=8192$, exceeding all baselines and reaching 95\% of T-optimal. More candidates cover the Co-SR configuration space more densely, making a near-optimal choice available. The learned surrogate improves to \SI{656}{\mega\bit\per\second} at $N=1024$ and plateaus beyond $N=2048$, while the upper bound continues to improve. Finding increasingly rare good configurations is thus the limiting factor. Nevertheless, it exceeds DCF by 16\% without any online interaction. H-MAB reaches higher average throughput but needs convergence over many steps. SR groups achieves a similar average, but its runs span \SI{473}{\mega\bit\per\second}--\SI{1205}{\mega\bit\per\second}, which limits scheduling predictability. It also requires cross-link RSSI measurements for every AP--STA pair, which are both unreliable and expensive to collect and exchange. In contrast, our method relies only on readily available measurements and incorporates signal strength as graph features, rather than making threshold-based decisions from individual RSSI values, which are inherently noisy and error-prone.

We next examine how well the surrogate generalizes to new topologies of varying size. Fig.~\ref{fig:boxplot} evaluates unseen open-space topologies with 2--16~APs, where each point aggregates 10 independently drawn layouts, using top-4 selection and an AP-dependent sample budget (since the configuration space grows with network size, we scale $N$ with the number of APs, from $N = 64$ candidates at 2--4~APs and doubling every two APs, up to $N = 512$ at 14--16~APs). The top-4 configurations form a pre-computed Co-SR catalog: when a sharing AP wins DCF contention and belongs to a stored group, the corresponding shared APs join the transmission. The reported metric covers Co-SR transmissions only. The surrogate recovers 52--77\% of the perfect surrogate median rate across all topology sizes, matching most closely at 2~APs (77\%) and dropping to 52--54\% for 14--16~APs, while maintaining consistently low variability across scenarios. The largest topologies are out-of-distribution, as training covers 2--10~APs.

\begin{table}[t]
    \caption{Surrogate computation time vs.\ H-MAB convergence time [\si{\milli\second}]. Dashes indicate out-of-memory conditions under full-batch inference.}
    \footnotesize
    \label{tab:timing}
    \centering
    \begin{tabular}{@{\hskip 4pt}c|rr|rr@{\hskip 4pt}}
        \toprule
        \multirow{2}{*}{\textbf{N}} & \multicolumn{2}{c|}{\textbf{Small (4 APs)}} & \multicolumn{2}{c}{\textbf{Large (16 APs)}} \\
        \cmidrule(lr){2-3} \cmidrule(l){4-5}
        & \textbf{Full} & \textbf{Split} & \textbf{Full} & \textbf{Split} \\ \midrule
        4    & 1.5   & 1.5   & 1.9   & 1.9   \\
        8    & 1.7   & 1.7   & 2.7   & 2.7   \\
        16   & 2.0   & 2.0   & 6.0   & 6.0   \\
        32   & 2.8   & 3.9   & 18.5  & 12.1  \\
        64   & 6.2   & 7.8   & 66.2  & 24.1  \\
        128  & 18.8  & 15.4  & 313.8 & 47.8  \\
        256  & 66.6  & 30.2  & --    & 95.2  \\
        512  & 314.5 & 60.4  & --    & 190.7 \\
        1024 & --    & 120.7 & --    & 381.1 \\ \midrule
        \multicolumn{1}{@{\hskip 4pt}c|}{H-MAB}  & \multicolumn{2}{c|}{3780} & \multicolumn{2}{c@{\hskip 4pt}}{\emph{inf}} \\
        \bottomrule
    \end{tabular}
\end{table}

Table~\ref{tab:timing} reports wall-clock inference times on a server with NVIDIA GH200 and \SI{118}{\giga\byte} of RAM, for a surrogate of ${\sim}1$M parameters. We compare two batching strategies: \emph{full batch} (all $N$ graphs in one pass) and \emph{split batch} (chunks of 16), where full-batch attention matrix scales quadratically with $N$, while split batching restores linear scaling in time. In the small scenario (4 APs), split batching scores 1024 candidates in \SI{121}{\milli\second}; in the large scenario (16 APs), the same workload takes \SI{381}{\milli\second}. Crucially, full-batch inference runs out of memory for $N > 128$. H-MAB requires \SI{3780}{\milli\second} to converge (i.e., reach its performance plateau) in the small scenario and fails entirely in the large one~\cite{wojnar2025ieee}. Our surrogate is over 30$\times$ faster with no online interaction. Additionally, as the surrogate produces a catalog that is reused across many subsequent TXOPs and refreshed periodically or upon a significant topology change, its inference cost is amortized. This makes it feasible to apply the method even in CPU-based deployments.

\section{Conclusion}
\label{sec:conclusion}

We have presented a surrogate-based approach to Co-SR optimization: draw random configurations, score them, and select the best. The perfect surrogate shows that this strategy reaches 95\% of T-optimal throughput by increasing catalog coverage. The surrogate model already exceeds DCF without online interaction and evaluates 1024 candidates over 30$\times$ faster than H-MAB convergence, while requiring only sparse, readily available RSSI links. Its plateau beyond 2048 candidates identifies ranking many configurations as the main limitation and the clearest direction for future work.

Several simplifications constrain the current evaluation: we do not use a full-stack simulator (e.g., ns-3), and only static topologies with full-buffer downlink traffic are considered. In addition, extending the method to other MAPC schemes (e.g., coordinated beamforming) and analyzing alternative scheduling strategies beyond top-$k$ are important directions for future work. Finally, a natural next step is to validate our approach using both a packet-level simulator and a hardware testbed.

\bibliographystyle{IEEEtran}
\bibliography{bibliography}

\end{document}